\documentclass[aps,prl,amsmath,twocolumn]{revtex4}

\preprint{JLAB-THY-14-1954}

\usepackage{epsfig}
\usepackage{graphics}
\usepackage{latexsym}
\usepackage{amsmath}
\usepackage{amssymb}
\usepackage{rotating}
\usepackage{subfigure}
\usepackage{bm}
\usepackage{color}

\begin{document}

\title{$\bar d - \bar u$ asymmetry in the proton in chiral effective theory}
\author{Yusupujiang Salamu$^{1}$}
\author{Chueng-Ryong Ji$^{2}$}
\author{W. Melnitchouk$^{3}$}
\author{P. Wang$^{1,4}$}
\affiliation{$^1$Institute of High Energy Physics, CAS,
		 Beijing 100049, China}
\affiliation{$^2$North Carolina State University,
		 Raleigh, North Carolina 27695, USA}
\affiliation{$^3$Jefferson Lab, Newport News,
		 Virginia 23606, USA}
\affiliation{$^4$Theoretical Physics Center for Science Facilities, CAS,
		 Beijing 100049, China}

\begin{abstract}
We compute the $\bar d - \bar u$ asymmetry in the proton in chiral
effective theory, including both nucleon and $\Delta$ degrees of
freedom, within the relativistic and heavy baryon frameworks.
In addition to the distribution at $x>0$, we compute the corrections
to the asymmetry from zero momentum contributions from pion rainbow
and bubble diagrams at $x=0$, which have not been accounted for in
previous analyses.
We find that the empirical $x$ dependence of $\bar d - \bar u$
as well as the integrated asymmetry can be well reproduced in
terms of a transverse momentum cutoff parameter.
\end{abstract}

\maketitle

The observation of the $\bar d-\bar u$ flavor asymmetry in the light
quark sea of the proton \cite{NMC, HERMES, NA51, E866} has been one
of the seminal results in hadronic physics over the past two decades,
leading to a major reevaluation of our understanding of the quark
structure of the nucleon.  In particular, the measurement revealed
the importance of 5-quark Fock state components of the nucleon's
wave function, and the crucial role played by chiral symmetry breaking.
This asymmetry had been anticipated by Thomas \cite{Thomas83}
a decade earlier, and has subsequently been studied using various
nonperturbative models
\cite{Signal91, Kumano98, Speth98, Chang11, Alberg12}.
However, despite some successes in reproducing the general features
of the data, it has proved very difficult to obtain direct connection
between the models and QCD.

An important development in establishing a formal link between
models of $\bar d-\bar u$ and QCD came with the realization that
the moments of parton distribution functions (PDFs) could be
formally expanded in chiral effective field theory in terms of
power series in the pion mass squared, $m_\pi^2$.
The leading nonanalytic (LNA) contributions were found to depend
on the (model-independent) long range structure of the pion cloud,
with a characteristic $m_\pi^2 \log m_\pi^2$ dependence
\cite{TMS00, Chen02}.
This idea was later applied to the chiral extrapolation of lattice
QCD moments of the $u-d$ distribution to reconcile the lattice
data at large $m_\pi^2$ with experiment \cite{Detmold01}.

Initial calculations of pion loop effects were performed in the
context of the ``Sullivan'' process \cite{Sullivan72}, using
pseudoscalar pion--nucleon coupling, which involves only the pion
``rainbow'' diagram.  Analysis within the chiral effective theory
for the pseudovector coupling reveals differences in the off-shell
behavior of the loops \cite{Comment13}, as well as the presence
of additional pion bubble terms \cite{Arndt01, Chen01} at $x=0$
\cite{BHJMT13}.
The relationship between the pseudoscalar and pseudovector
theories was recently discussed in Refs.~\cite{BHJMT13, JMT13}.

While the structure of the pion loops constrains the behavior of
the PDF moments at small $m_\pi^2$, the total moments depend also
on the short-distance contributions, parametrized by
coefficients of analytic terms in the chiral expansion.
In principle, these can be fitted to data, and the PDFs reconstructed
from the moments assuming a functional form for the dependence on the
parton momentum fraction $x$ \cite{Detmold-x}.
Recently, however, a method for computing the PDFs directly in
$x$ space has been developed, by matching nonlocal operators
within heavy baryon (HB) chiral effective theory \cite{Moiseeva13}.
The results allow the total distributions to be computed in the
form of convolutions of bare PDFs in the pion with pion light-cone
momentum distributions associated with pion rainbow, bubble,
and Kroll-Ruderman terms \cite{BHJMT13}.

In this letter, we apply this formalism for the first time to
analyze the $\bar d-\bar u$ distribution in the proton within
chiral effective theory.  We consider both the covariant and
nonrelativistic HB formulations of the low-energy chiral theory,
including both nucleon $N$ and $\Delta$ contributions to loop
integrals.
While the on-shell components of the $N$ and $\Delta$ rainbow
diagrams give rise to distributions that closely resemble
earlier model calculations, we find in addition unique
signatures of off-shell components and bubble diagrams that
are nonzero at $x=0$.
Although difficult to access directly through experiment,
these terms modify the value of $\bar d-\bar u$ integrated
over all $x$, which is the benchmark for the magnitude of the
flavor symmetry violation.
To illustrate the phenomenological application of our approach,
we also compute the shape of the $\bar d-\bar u$ distribution
using a transverse momentum cutoff to regularize the ultraviolet
contributions from the pion loops.

\begin{figure}[t]
\includegraphics[width=8cm]{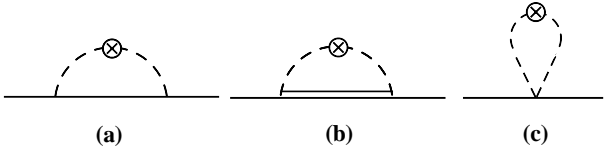}
\caption{Contributions to the pion light-cone momentum distributions
	in the proton, from the pion rainbow diagram with (a) a nucleon
	or (b) $\Delta$ intermediate state, and (c) from the pion
	bubble diagram.}
\label{fig:loops}
\end{figure}

The pion light-cone distributions can be calculated in the effective
chiral theory from the diagrams shown in Fig.~\ref{fig:loops}
\cite{BHJMT13, JMT13, Moiseeva13}.  Other contributions, such as
the rainbow diagrams with coupling to the baryon, pion tadpoles,
or Kroll-Ruderman diagrams, also give nonzero corrections to PDFs,
but do not contribute to $\bar d-\bar u$ if one assumes a flavor
symmetric sea in the bare proton.

Starting from the lowest-order chiral Lagrangian
\cite{Jenkins91, Bernard08},
these have previously been derived for the diagrams in
Fig.~\ref{fig:loops}(a) and (c) involving nucleons and pions
\cite{BHJMT13, Moiseeva13}.
Including also contributions from the $\Delta$ intermediate states
in Fig.~\ref{fig:loops}(b), the $\bar d-\bar u$ difference in the
proton can be written as \cite{Chen02}
\begin{equation}
\bar d - \bar u
= \left( f_{\pi^+ n} + f_{\pi^+ \Delta^0} - f_{\pi^-\Delta^{++}}
       + f_{\pi (\rm bub)}
  \right) \otimes \bar{q}_v^\pi,
\label{eq:conv}
\end{equation}
where generally the convolution is defined as
$f \otimes q = \int_0^1 dy \int_0^1 dz\, \delta(x-yz)\, f(y)\, q(z)$,
with $y=k^+/p^+$ the light-cone fraction of the proton's
momentum ($p$) carried by the pion ($k$).
The pion light-cone momentum distributions $f_{\pi N}$ and
$f_{\pi \Delta}$ correspond to the pion rainbow diagrams in
Figs.~\ref{fig:loops}(a) and (b), while $f_{\pi (\rm bub)}$
represents the pion bubble diagram in Fig.~\ref{fig:loops}(c).
The contributions from individual charge states in Eq.~(\ref{eq:conv})
arise from the fluctuations
$p \to \pi^+\, n$, $\pi^+\, \Delta^0$ or $\pi^-\, \Delta^{++}$.
%
The convolution in Eq.~(\ref{eq:conv}) is obtained from the
crossing symmetry properties of the light-cone distributions
\cite{Chen02}, $f(-y) = f(y)$, and the valence pion PDF,
$\bar{q}^\pi(x) = -q^\pi(-x)$, for which we have assumed
charge symmetry,
$\bar{q}_v^\pi
\equiv \bar{d}^{\pi^+} - d^{\pi^+}
     = \bar{u}^{\pi^-} - u^{\pi^-}$.

Following \cite{BHJMT13}, the $f_{\pi^+ n}$ distribution
can be written as
\begin{equation}
f_{\pi^+ n}(y)
= 2 \Big[ f_N^{\rm (on)}(y) + f_N^{\rm (\delta)}(y) \Big],
\end{equation}
where $f_N^{\rm (on)}$ and $f_N^{\rm (\delta)}$ are the on-shell
and $\delta$-function contributions from the pion rainbow diagram,
respectively.  The on-shell nucleon term for $y > 0$ is
\cite{Thomas83, Kumano98, Speth98, BHJMT13}
\begin{equation}
f_N^{\rm (on)}(y)
= \frac{g_A^2 M^2}{(4\pi f_\pi)^2}
  \int\!dk_\bot^2\,
  \frac{y\, (k_\bot^2 + y^2 M^2)}{(1-y)^2 D_{\pi N}^2},
\label{eq:fNon}
\end{equation}
where $M$ is the nucleon mass,
$g_A = 1.267$ is the axial charge,
$f_\pi = 93$~MeV is the pion decay constant, and
$D_{\pi N} = -[k_\bot^2 + y^2 M^2 + (1-y) m_\pi^2]/(1-y)$
is the pion virtuality ($k^2-m_\pi^2$) for an on-shell
nucleon intermediate state.
In contrast, the off-shell contribution arises from
pions with zero light-cone momentum,
\begin{equation}
f_N^{(\delta)}(y) 
= {g_A^2 \over 4 (4\pi f_\pi)^2} \int\!dk_\perp^2\,
  \log\frac{\Omega_\pi}{\mu^2}\,
  \delta(y),
\label{eq:fydelta}
\end{equation}      
where $\Omega_\pi = k_\bot^2 + m_\pi^2$,
and $\mu$ is an ultraviolet cutoff on the $k^-$ integration.
Since $f^{(\delta)}$ is nonzero only at $y=0$, it contributes
to $\bar d-\bar u$ only at $x=0$.
While it cannot be measured directly, it nevertheless
affects the determinaton of
  $\overline{D} - \overline{U} \equiv \int_0^1 dx (\bar d - \bar u)$
(as in the Gottfried sum rule \cite{Gottfried67}) when extrapolating
to $x=0$.

The $\pi\Delta$ contribution can be computed from the effective
$\pi N \Delta$ interaction
$\bar\psi_\Delta^\mu
 [g_{\mu\nu} - (Z+1/2) \gamma_\mu \gamma_\nu]
 \partial^\nu\phi_\pi\, \psi_N$,
where $Z$ parametrizes the off-shell behavior of the
spin-3/2 field $\psi_\Delta^\mu$.
The resulting $\pi\Delta$ distribution function can be
written as a sum of three terms,
\begin{equation}
f_{\pi^+ \Delta^0}(y)
= f_\Delta^{\rm (on)}(y)
+ f_\Delta^{\rm (end\textrm{-}pt)}(y)
+ f_\Delta^{(\delta)}(y).
\label{eq:fypiD}
\end{equation}
For $Z=-1/2$, the on-shell ($\Delta$-pole) contribution at
\mbox{$y > 0$} is given by
\begin{eqnarray}
f_\Delta^{\rm (on)}(y)
&=& C_\Delta \int\!dk_\bot^2\,
  \frac{y\, (\overline{M}^2-m_\pi^2)}{1-y}	\nonumber\\
& & \hspace*{-2cm} \times
  \Big[
    \frac{(\overline{M}^2\!-\!m_\pi^2)(\Delta^2\!-\!m_\pi^2)}
	 {D_{\pi \Delta}^2}
  - \frac{3(\Delta^2\!-\!m_\pi^2) + 4M M_\Delta}
	 {D_{\pi \Delta}}
  \Big]
\label{eq:fDon}
\end{eqnarray}
where
\mbox{$D_{\pi\Delta}
= -[k_\bot^2 - y(1-y) M^2 + y M_\Delta^2 + (1-y) m_\pi^2]/$}
\mbox{$(1-y)$}
is the pion virtuality for an on-shell $\Delta$ intermediate state,
and we define
  $\Delta \equiv M_\Delta - M$ and
  $\overline{M} \equiv M_\Delta + M$.
The coefficient
$C_\Delta = g_{\pi N\Delta}^2 / [(4\pi)^2 18 M_\Delta^2]$,
and using SU(6) symmetry and the Gell-Mann--Oakes--Renner relation
the $\pi N \Delta$ coupling constant is
  $g_{\pi N\Delta}
  = (3\sqrt{2}/5) g_A/f_\pi
  \approx 11.8$~GeV$^{-1}$ \cite{MST99}.

In addition, the pole contribution involves an end-point singularity,
which gives a $\delta$-function at $y=1$,
\begin{eqnarray}
f_\Delta^{\rm (end\textrm{-}pt)}(y)
&=& C_\Delta \int\!dk_\bot^2\, \delta(1-y) 		\nonumber\\
& & \hspace*{-2.3cm} \times
  \Big\{
     \Big[ \Omega_\Delta - 2(\Delta^2 - m_\pi^2) - 6 M M_\Delta \Big]
     \log\frac{\Omega_\Delta}{\mu^2}	
   - \Omega_\Delta
  \Big\}
\label{eq:fDend}
\end{eqnarray}
where $\Omega_\Delta = k_\bot^2 + M_\Delta^2$.
The off-shell components of the $\Delta$ propagator introduce
a $\delta$-function term at $y=0$,
\begin{eqnarray}
f_\Delta^{(\delta)}(y)
&=& C_\Delta \int\!dk_\bot^2\, \delta(y)		\nonumber\\
& & \hspace*{-1.5cm} \times
  \Big\{
     \Big[ 3 (\Omega_\pi + m_\pi^2) + \overline{M}^2 \Big]
     \log\frac{\Omega_\pi}{\mu^2}
   - 3 \Omega_\pi			
  \Big\}.
\label{eq:fDdel}
\end{eqnarray}
Although this gives a nonzero PDF only at $x = 0$, since it
contributes to the integral of $\bar d - \bar u$, it will
indirectly affect the normalization for $x > 0$.
For the $p \to \Delta^{++} \pi^-$ dissociation,
the distribution function is given by
$f_{\pi^- \Delta^{++}} = 3 f_{\pi^+ \Delta^0}$.
Note also that for values of the off-shell parameter $Z \neq -1/2$,
the additional interaction term $\sim \gamma_\mu \gamma_\nu$
contributes only to the off-shell component $f_\Delta^{(\delta)}$,
without modifying our LNA result.

The $\Delta$ contribution has been considered in several previous
studies, both within pion cloud models \cite{Speth98, Kumano98,
Alberg12, MST99} and in chiral effective theory in the large-$N_c$
limit \cite{Chen_Delta}.
In the phenomenological approaches, one computes the ``Sullivan''
process with the $\Delta$ intermediate state by taking the
$\Delta$-pole contribution, $(p-k)^2 - M_\Delta^2 \to 0$,
which gives the distribution usually found in the literature
\cite{Kumano98, Speth98, Alberg12, TMS00, MST99},
\begin{eqnarray}
f_\Delta^{\rm (Sul)}(y)
&=& C_\Delta \int\!dk_\bot^2\, y		\nonumber\\
& & \hspace*{-1.5cm} \times
    \frac{\big[ k_\bot^2 + (\Delta+y M)^2 \big]
	  \big[ k_\bot^2 + (\overline{M}-y M)^2 \big]^2}
         {(1-y)^4 D_{\pi \Delta}^2}.
\label{eq:fDsul}
\end{eqnarray}
Note that the power of $k_\bot$ in the numerator here is $k_\bot^6$,
while in the on-shell contribution in Eq.~(\ref{eq:fDon}) it is
$k_\bot^2$.  This difference arises because the Sullivan process
neglects the end-point contributions, which give rise to the
$\delta(1-y)$ term in Eq.~(\ref{eq:fDend}), and also cancel the
${\cal O}(k_\bot^4)$ and ${\cal O}(k_\bot^6)$ terms.
The correct calculation of the $\Delta$-pole contribution therefore
yields $f_\Delta^{\rm (on)} + f_\Delta^{\rm (end\textrm{-}pt)}$.
The off-shell contribution proportional to $\delta(y)$ is not
included in the Sullivan approach, for either the $\Delta$ or
nucleon intermediate states.  For the latter, the Sullivan
method yields only the on-shell component \cite{BHJMT13, JMT13},
$f_N^{\rm (Sul)}(y) = f_N^{\rm (on)}(y)$.
%

Finally, for the pion bubble diagram in Fig.~\ref{fig:loops}(c),
the distribution $f_{\pi (\rm bub)}$ has a form similar to the
$\delta$-function part of $f_{\pi^+ n}$ \cite{BHJMT13},
\begin{equation}
f_{\pi (\rm bub)}(y)
= -{2 \over g_A^2}\, f_N^{(\delta)}(y).
\label{eq:fybub}
\end{equation}      
This term originates with the Weinberg-Tomozawa part
of the chiral Lagrangian, and is independent of $g_A$.

Because the leading chiral behavior of matrix elements is not
affected by baryon masses, for simplicity most earlier studies of
chiral corrections to PDF moments \cite{Chen02, Arndt01, Chen01},
as well as generalized parton distributions \cite{Wang10},
were performed in the HB limit.
It is instructive therefore to compare the results for the $x$
dependence of PDFs in the HB and relativistic approaches
\cite{JMT13, Dorati08}.

In the HB limit ($m_\pi \ll M$, $y \ll 1$), the nonrelativistic
analog of the on-shell function in Eq.~(\ref{eq:fNon}) is obtained
by the replacement
$(1-y) D_{\pi N} \to \widetilde{D}_{\pi N}
		   = -(k_\bot^2 + y^2 M^2 + m_\pi^2)$.
The $\delta$-function and bubble contributions $f_N^{(\delta)}$
and $f_{\pi (\rm bub)}$, however, remain unchanged.

For the $\Delta$ intermediate state, in the HB limit both the $N$
and $\Delta$ masses are large, $\overline{M} \to \infty$, while
the difference is kept finite, $\Delta/\overline{M} \to 0$.
In this case the on-shell function reduces to
\begin{equation}
\widetilde{f}_{\Delta}^{\rm (on)}(y)
= \frac{8 g_{\pi N \Delta}^2 M^2}{9 (4\pi)^2}\!
  \int\!dk_\bot^2\,
  \frac{y \big(\Delta^2\!-\!m_\pi^2\!-\!\widetilde{D}_{\pi\Delta}\big)}
       {\widetilde{D}^2_{\pi\Delta}},
\label{eq:fDonNR}
\end{equation}
where the nonrelativistic analog of $D_{\pi\Delta}$ in
Eq.~(\ref{eq:fDon}) is
$\widetilde{D}_{\pi\Delta}
= -(k_\bot^2 + y^2 M^2 + 2y M \Delta + m_\pi^2)$.
Similarly, in the HB limit the $\delta$-function contribution
in Eq.~(\ref{eq:fDdel}) becomes
\begin{equation}
\widetilde{f}_\Delta^{(\delta)}(y)
= \frac{2 g_{\pi N \Delta}^2}{9 (4\pi)^2}\!
  \int\!dk_\bot^2\, \delta(y)
  \log\frac{\Omega_\pi}{\mu^2}.
\label{eq:fDdelNR}
\end{equation}
Because in this limit one has $y \ll 1$, there is no analogous
nonrelativistic end-point contribution to that in Eq.~(\ref{eq:fDend}).

The consistency of the above results with the chiral symmetry of
QCD can be verified by examining the LNA behavior of
$\overline{D} - \overline{U}$.  Since the valence pion PDF
$\bar q_v^\pi$ is normalized to unity, $\overline{D} - \overline{U}$
is given entirely by the moments of the pion distribution functions
in Eq.~(\ref{eq:conv}).  Expanding these in $m_\pi$, the LNA behavior
of $\overline{D} - \overline{U}$ is then given by
\begin{equation}
(\overline{D} - \overline{U})_{_{\rm LNA}}
= \frac{3g_A^2+1}{2(4\pi f_\pi)^2} m_\pi^2 \log m_\pi^2
- \frac{g_{\pi N\Delta}^2}{12\pi^2} J_1,
\label{eq:LNA}
\end{equation}
where
$J_1 = (m_\pi^2 - 2 \Delta^2) \log m_\pi^2
     + 2 \Delta\, r\, \log[(\Delta-r)/(\Delta+r)]$,
with $r = \sqrt{\Delta^2-m_\pi^2}$.
The nucleon intermediate state contribution in Eq.~(\ref{eq:LNA})
coincides with the result obtained in Ref.~\cite{Chen02}, and the
expression for $J_1$ agrees with that in Ref.~\cite{Arndt01}.
Note that the end-point component $f_\Delta^{\rm (end\textrm{-}pt)}$
does not contain any nonanalytic structure in $m_\pi^2$,
and therefore does not contribute to the LNA behavior.
Since the chiral properties of the pion distributions are
independent of the short-distance part of the pion--nucleon
interaction, the LNA contribution computed in the HB limit
is identical to that in Eq.~(\ref{eq:LNA}).
In the $\Delta \to 0$ limit, the coefficient of the LNA
$m_\pi^2 \log m_\pi^2$ term is 
$[(27/50) g_A^2 + 1/2]/(4\pi f_\pi)^2$.
Compared to the coefficient of the LNA term in Eq.~(\ref{eq:LNA})
from the nucleon alone, the $\Delta$ intermediate state leads to
a reduction in the LNA coefficient by more than 50\%.

On the other hand, in the Sullivan approach, which only accounts
for the on-shell components, the LNA behavior in the
$\Delta \to 0$ limit is \cite{TMS00}
\begin{equation}
(\overline{D} - \overline{U})_{_{\rm LNA}}^{\rm (Sul)}
= \left[ \frac{2 g_A^2}{(4\pi f_\pi)^2}
       - \frac{g_{\pi N \Delta}^2}{9\pi^2}
  \right]
  m_\pi^2 \log m_\pi^2.
\label{eq:LNAsul}
\end{equation}
For the nucleon rainbow contribution in Fig.~\ref{fig:loops}(a),
the on-shell (Sullivan) approximation is therefore a factor of
4/3 larger than the exact result in Eq.~(\ref{eq:LNA}).
Interestingly, the contribution from the $\Delta$ rainbow diagram
in Fig.~\ref{fig:loops}(b) in the Sullivan approach is also 4/3
larger than the full expression in the $\Delta \to 0$ limit
with the $\delta$-function components.
Using SU(6) couplings, the coefficient of the LNA term in the
Sullivan approximation is $(18/25) g_A^2/(4\pi f_\pi)^2$.

\begin{figure}[t]
\centering
{\vspace*{-2cm}\includegraphics[width=9.5cm]{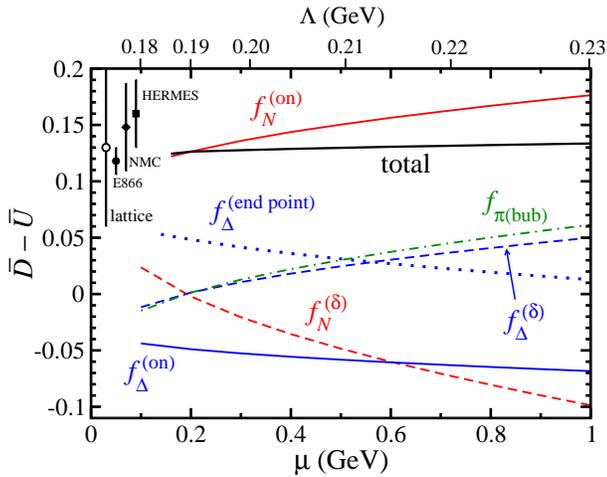}}
\caption{Contributions to $\overline{D}-\overline{U}$ from
	the diagrams in Fig.~\ref{fig:loops} as a function
	of the cutoff parameters $\mu$ and $\Lambda$.
	The nucleon on-shell and $\delta$-function,
	and $\Delta$ on-shell, end-point and $\delta$-function
	contributions are shown individually, together with the
	pion bubble contribution and the total.
	The calculations are compared with data from the
	NMC \cite{NMC}, 		
	HERMES \cite{HERMES} and 	
	E866 \cite{E866}		
	experiments, and from a recent lattice QCD calculation
	\cite{Lin14}.}			
\label{fig:comp}
\end{figure}

In Fig.~\ref{fig:comp} we show the individual contributions to
$\overline{D} - \overline{U}$ from the nucleon, $\Delta$ and
bubble diagrams as a function of $\mu$ and the $k_\bot$ cutoff
parameter $\Lambda$.
%
%
For $\mu$ ranging between 0.1 and 1~GeV, $\Lambda$ is fixed by
matching the calculation to the value of the $\bar d-\bar u$ integral
extracted from the E866 Drell-Yan data over the measured $x$ range,
$\int_{0.015}^{0.35} dx (\bar d-\bar u) = 0.0803(11)$ \cite{E866}.
The resulting variation in $\Lambda$ is relatively mild, ranging
from $\Lambda \approx 0.18$ to 0.23~GeV, and the total
$\overline{D} - \overline{U}$ is similar to the empirical
results from the Drell-Yan \cite{E866} and deep-inelastic
scattering \cite{NMC, HERMES} data.
One should caution, however, that the experimental values
are obtained by extrapolting the data to $x=0$ and $x=1$
under the assumption that there is no contribution at $x=0$.
The $\delta$-function contributions $f_N^{(\delta)}$,
$f_\Delta^{(\delta)}$ and $f_{\pi\rm (bub)}$ will give
nonzero corrections to the extrapolated moment.

Numerically, the most important contribution is from the nucleon
on-shell distribution $f_N^{(\rm on)}$, but this is cancelled to
some extent by the negative $f_N^{(\delta)}$ distribution.
The on-shell $\Delta$ contribution $f_\Delta^{(\rm on)}$ is negative
and $\approx 1/2$ the magnitude of the on-shell $N$ component.
On the other hand, the $\Delta$ contributions at $y=0$ and $y=1$
are both positive, with the sum of the two largely canceling the
negative on-shell $\Delta$ term.
The net result is a significantly smaller $\Delta$ component
than that found in many previous analyses.
The pion bubble contribution is positive for most values of
$\mu$ and slightly enhances the on-shell nucleon term.
From Eq.~(\ref{eq:fybub}) one expects the $f_N^{(\delta)}$
contribution to cancel strongly with $f_{\pi (\rm bub)}$,
and partially with the positive $f_\Delta^{(\delta)}$.
Thus we see strong cancellations between all the singular
$x=0$ pieces, leaving the total pionic contributions to
$\overline{D} - \overline{U}$ determined largely by the
nucleon on-shell part \cite{Thomas83}.
This explains for the first time the relative success of the
phenomenological descriptions of the data through the Sullivan
process in terms of on-shell nucleon contribution alone.
In practice, the $\delta$-function pieces at $x=0$ can contribute
up to $\approx 16\%$ (for $\mu=1$~GeV) of the $\bar d-\bar u$
difference integrated over the measured region of $x$, which is
a relatively small contribution to the Gottfried sum violation.

For the nonrelativistic calculation, with the same values of the
cutoff parameters, there is a small reduction in the $N$ contribution,
reflecting the $D_{\pi N} \to \widetilde{D}_{\pi N}$ modification
in the on-shell component.
On the other hand, the absence of the end-point term in the
HB calculation means that the nonrelativistic on-shell
contribution is significantly more negative, and cancels much more
of the total nucleon contribution.

\begin{figure}[t]
\centering
\includegraphics[width=8cm]{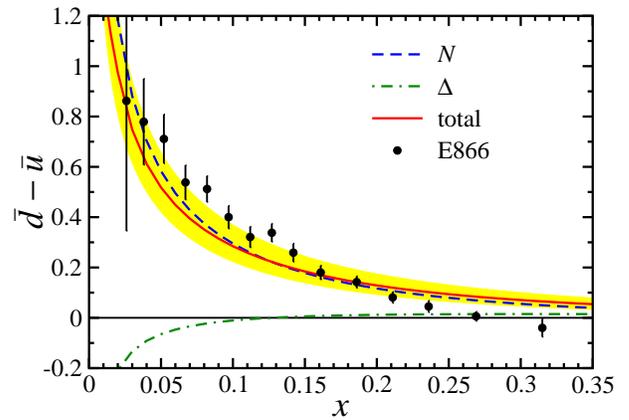}
\caption{Flavor asymmetry $\bar d - \bar u$ from the $N$ and
	$\Delta$ intermediate states, and the total, for cutoffs
	$\mu=0.3$~GeV and $\Lambda=0.2$~GeV, compared with the
	asymmetry extracted at leading order from the E866 Drell-Yan
	data \cite{E866} at $Q^2=54$~GeV$^2$.  The band indicates
	the uncertainty on the total distribution from the cutoff
	parameters (for $\mu$ between 0.1 and 1~GeV) and from the
	empirical $\overline{D} - \overline{U}$	normalization.}
\label{fig:dbar-ubar}
\end{figure}

Within the framework of the present calculation, we can also estimate
the $x$ dependence of $\bar d - \bar u$ from the convolution in
Eq.~(\ref{eq:conv}), with the light-cone distributions computed
in terms of the same parameters as in Fig.~\ref{fig:comp}.
For the valence PDFs in the pion we use the global parametrization
from Ref.~\cite{GRS99}.
The resulting $\bar d - \bar u$ asymmetry is illustrated in
Fig.~\ref{fig:dbar-ubar} at a scale $Q^2=54$~GeV$^2$, with the
nucleon and $\Delta$ contributions at $x > 0$ computed for
$\mu=0.3$~GeV and $\Lambda=0.2$~GeV (the $\delta$-function
contributions $f_N^{(\delta)}$, $f_\Delta^{(\delta)}$ and
$f_{\pi(\rm bub)}$ exist only at $x=0$).
The asymmetry is dominated by the nucleon on-shell component,
with the $\Delta$ on-shell contribution providing some cancellation
at small $x$, but becoming negligible for $x \gtrsim 0.1$.
The shaded band illustrates the uncertainty in the calculation,
with the envelope representing the extremal values of the
cutoffs $\mu=0.1$~GeV and 1~GeV, and the experimental
uncertainty on the E866 data \cite{E866}.
Without attempting to fine-tune the parameters, the overall
agreement between the calculation and experiment is very good.
As with all previous pion loop calculations, the apparent trend
of the E866 data towards negative $\bar d - \bar u$ values for
$x \gtrsim 0.3$ is not reproduced in this analysis.
The new SeaQuest experiment \cite{E906} at Fermilab is expected
to provide new information on the shape of $\bar d - \bar u$
for $x \lesssim 0.45$.

%

The analysis described here can be applied to other
nonperturbative quantities in the proton, such as the flavor
asymmetry of the polarized sea, $\Delta\bar u - \Delta\bar d$,
or the strange--antistrange asymmetry $s - \bar s$ using an SU(3)
generalization of the effective chiral theory.  Beyond this,
the methodology can be further extended to study the systematics
of chiral loop corrections to partonic observables such as
transverse momentum dependent distributions and generalized
parton distributions.

We are grateful to A.W.~Thomas for helpful comments.
This work was supported by the DOE Contract No.~DE-AC05-06OR23177,
under which Jefferson Science Associates, LLC operates Jefferson Lab,
DOE Contract No.~DE-FG02-03ER41260, and by NSFC under Grant
No.~11261130311 (CRC 110 by DFG and NSFC).



\begin{thebibliography}{99}

\bibitem{NMC}
M.~Arneodo {\it et al.},
Phys. Rev. D {\bf 50}, 1 (1994).

\bibitem{HERMES}
K.~Ackerstaff {\it et al.},
Phys. Rev. Lett. {\bf 81}, 5519 (1998).

\bibitem{NA51}
A.~Baldit {\it et al.},
Phys. Lett. B {\bf 332}, 244 (1994).

\bibitem{E866}
R.~S.~Towell {\it et al.},
Phys. Rev. D {\bf 64}, 052002 (2001).

\bibitem{Thomas83}
A.~W.~Thomas,
Phys. Lett. B {\bf 126}, 97 (1983).

\bibitem{Signal91}
A.~I.~Signal, A.~W.~Schreiber and A.~W.~Thomas,
Mod. Phys. Lett. A {\bf 6}, 271 (1991).

\bibitem{Kumano98}
S.~Kumano,
Phys. Rep. {\bf 303}, 183 (1998).

\bibitem{Speth98}
J.~Speth and A.~W.~Thomas,
Adv. Nucl. Phys. {\bf 24}, 83 (1998).

\bibitem{Chang11}
W.-C.~Chang and J.-C.~Peng,       
Phys. Rev. Lett. {\bf 106}, 252002 (2011);
%
J.-C.~Peng and J.-W.~Qiu,
Prog. Part. Nucl. Phys. {\bf 76}, 43 (2014).

\bibitem{Alberg12}
M.~Alberg and G.~A.~Miller,
Phys. Rev. Lett. {\bf 108}, 172001 (2012).

\bibitem{TMS00}
A.~W.~Thomas, W.~Melnitchouk and F.~M.~Steffens,
Phys. Rev. Lett. {\bf 85}, 2892 (2000).

\bibitem{Chen02}
J.-W.~Chen and X.~Ji,
Phys. Rev. Lett. {\bf 87}, 152002 (2001);
{\bf 88}, 249901(E) (2002).

\bibitem{Detmold01}
W.~Melnitchouk, J.~W.~Negele, D.~B.~Renner and A.~W.~Thomas,
Phys. Rev. Lett. {\bf 87}, 172001 (2001).

\bibitem{Sullivan72}
J.~D.~Sullivan,
Phys. Rev. D {\bf 5}, 1732 (1972).

\bibitem{Comment13}
C.-R.~Ji, W.~Melnitchouk and A.~W.~Thomas,
Phys. Rev. Lett. {\bf 110}, 179101 (2013),

\bibitem{Arndt01}
D.~Arndt and M.~J.~Savage,
Nucl. Phys. {\bf A697}, 429 (2002).

\bibitem{Chen01}
J.~W.~Chen and X.~Ji,
Phys. Lett. B {\bf 523}, 107 (2001).

\bibitem{BHJMT13}
M.~Burkardt, K.~S.~Hendricks, C.-R.~Ji, W.~Melnitchouk and A.~W.~Thomas,
Phys. Rev. D {\bf 87}, 056009 (2013).

\bibitem{JMT13}
C.-R.~Ji, W.~Melnitchouk and A.~W.~Thomas,
Phys. Rev. D {\bf 88}, 076005 (2013).

\bibitem{Detmold-x} 
W.~Detmold, W.~Melnitchouk and A.~W.~Thomas,
Eur. Phys. J. direct C {\bf 3}, 1 (2001).

\bibitem{Moiseeva13}
A.~M.~Moiseeva and A.~A.~Vladimirov,
Eur. Phys. J. A {\bf 49}, 23 (2013).

\bibitem{Jenkins91}
E.~E.~Jenkins and A.~V.~Manohar,
Phys. Lett. B {\bf 255}, 558 (1991).

\bibitem{Bernard08}
V.~Bernard,
Prog. Part. Nucl. Phys. {\bf 60}, 82 (2008).

\bibitem{Gottfried67}
K.~Gottfried,
Phys. Rev. Lett. {\bf 18}, 1174 (1967).

\bibitem{MST99}
W.~Melnitchouk, J.~Speth and A.~W.~Thomas,
Phys. Rev. D {\bf 59}, 014033 (1999).

\bibitem{Chen_Delta}
J.-W.~Chen and X.~Ji,
Phys. Lett. B {\bf 523}, 73 (2001)

\bibitem{Wang10}
P.~Wang and A.~W.~Thomas,
Phys. Rev. D {\bf 81}, 114015 (2010).

\bibitem{Dorati08}
M.~Dorati, T.~A.~Gail and T.~R.~Hemmert,
Nucl. Phys. {\bf A798}, 96 (2008).

\bibitem{Lin14}
H.-W.~Lin, J.-W.~Chen, S.~D.~Cohen and X.~Ji,
arXiv:1402.1462 [hep-ph].

\bibitem{GRS99}
M.~Gluck, E.~Reya and I.~Schienbein,
Eur. Phys. J. C {\bf 10}, 313 (1999).


\bibitem{E906}
Fermilab E906 experiment (SeaQuest),
D.~F.~Geesaman and P.~E.~Reimer, spokespersons.

\end{thebibliography}
\end{document}